\documentclass[sigconf, nonacm, pdfa]{acmart}
\usepackage{subcaption}
\usepackage{makecell}
\usepackage{multirow}
\usepackage{listings}
\usepackage{balance}
\usepackage{tabularx}
\usepackage{array}
\usepackage{xspace}
\usepackage[a-2b]{pdfx}

\newcommand\vldbdoi{10.14778/3705829.3705843}
\newcommand\vldbpages{253 - 264}
\newcommand\vldbvolume{18}
\newcommand\vldbissue{2}
\newcommand\vldbyear{2024}
\newcommand\vldbauthors{\authors}
\newcommand\vldbtitle{\shorttitle} 
\newcommand\vldbavailabilityurl{https://github.com/uiuc-kang-lab/leap}
\newcommand\vldbpagestyle{empty} 

\newcommand{\sysname}{\textsc{LEAP}\xspace}
\newcommand{\datasetname}{\textsc{QUIET-ML}\xspace}
\newcommand{\minihead}[1]{{\vspace{.5em}\noindent\textbf{#1.} }}

\lstdefinelanguage{CustomSQL}{
  basicstyle=\ttfamily\small,
  keywordstyle=\color{blue},
  morekeywords={SELECT, FROM, WHERE, GROUP, BY, COUNT, AS, LIKE},
  sensitive=true,
  commentstyle=\color{gray},
  stringstyle=\color{red},
  showstringspaces=false,
  moredelim=[is][\textcolor{red}]{@}{@},
}

\lstdefinestyle{CustomPython}{
  language=Python,
  basicstyle=\ttfamily\small,
  keywordstyle=\color{red},
  commentstyle=\color{gray},
  stringstyle=\color{red},
  showstringspaces=false,
  postbreak=\mbox{\textcolor{red}{$\hookrightarrow$}\space},
}

\begin{document}

\title{\sysname: LLM-powered End-to-end Automatic Library for Processing Social Science Queries on Unstructured Data}

\author{Chuxuan Hu}
\affiliation{%
  \institution{UIUC}
  \streetaddress{streetaddress}
  \city{Urbana}
  \state{IL}
  \postcode{postcode}
}
\email{chuxuan3@illinois.edu}

\author{Austin Peters}
\affiliation{%
  \institution{University of Chicago}
  \streetaddress{streetaddress}
  \city{Chicago}
  \state{IL}
  \postcode{postcode}
}
\email{austinpeters@uchicago.edu}

\author{Daniel Kang}
\affiliation{%
  \institution{UIUC}
  \streetaddress{streetaddress}
  \city{Urbana}
  \state{IL}
  \postcode{postcode}
}
\email{ddkang@illinois.edu}

\begin{abstract}

Social scientists are increasingly interested in analyzing the semantic information
(e.g., emotion) of unstructured data (e.g., Tweets), where the semantic
information is not natively present. Performing this analysis in a
cost-efficient manner requires using machine learning (ML) models to extract
the semantic information and subsequently analyze the now structured data.
However, this process remains challenging for domain experts.

To demonstrate the challenges in social science analytics, we collect a dataset, \datasetname, of 120 real-world social science queries in natural language and
their ground truth answers. Existing systems struggle with these queries since
(1) they require selecting and applying ML models, and (2) more than a quarter
of these queries are vague, making standard tools like natural language to SQL
systems unsuited. To address these issues, we develop \sysname, an end-to-end library
that answers social science queries in natural language with ML. \sysname filters vague queries to ensure that the answers are deterministic and selects from internally supported and user-defined ML functions to extend the unstructured data to structured tables with necessary annotations. \sysname further generates and executes code to respond to these natural language queries. \sysname achieves a 100\% pass @ 3 and 92\% pass @ 1 on \datasetname, with a
\$1.06 average end-to-end cost, of which code generation costs \$0.02.

\end{abstract}

\maketitle

\pagestyle{\vldbpagestyle}
\begingroup\small\noindent\raggedright\textbf{PVLDB Reference Format:}\\
\vldbauthors. \vldbtitle. PVLDB, \vldbvolume(\vldbissue): \vldbpages, \vldbyear.\\
\href{https://doi.org/\vldbdoi}{doi:\vldbdoi}
\endgroup
\begingroup
\renewcommand\thefootnote{}\footnote{\noindent
This work is licensed under the Creative Commons BY-NC-ND 4.0 International License. Visit \url{https://creativecommons.org/licenses/by-nc-nd/4.0/} to view a copy of this license. For any use beyond those covered by this license, obtain permission by emailing \href{mailto:info@vldb.org}{info@vldb.org}. Copyright is held by the owner/author(s). Publication rights licensed to the VLDB Endowment. \\
\raggedright Proceedings of the VLDB Endowment, Vol. \vldbvolume, No. \vldbissue\ %
ISSN 2150-8097. \\
\href{https://doi.org/\vldbdoi}{doi:\vldbdoi} \\
}\addtocounter{footnote}{-1}\endgroup

\ifdefempty{\vldbavailabilityurl}{}{
\vspace{.3cm}
\begingroup\small\noindent\raggedright\textbf{PVLDB Artifact Availability:}\\
The source code, data, and/or other artifacts have been made available at \url{\vldbavailabilityurl}.
\endgroup
}

\section{Introduction}
\label{sec:intro}

With increasingly accessible \emph{unstructured} datasets \cite{kang2017noscope, kang2022tasti, chu2007relational, wei2020analytics}, social scientists are able to answer questions that were previously beyond scope over unstructured data \cite{mcfarland2016sociology, annurev-soc-121919-054621}, 
ranging from
macroeconomic questions like ``Is the public mood correlated or even
predictive of economic indicators?'' \cite{BOLLEN20111} to sociolinguistic
questions like ``How can conversational behavior reveal power
relationships?'' \cite{Danescu-Niculescu-Mizil+al:12a}. However, domain experts face two major challenges in answering such queries.

The first challenge is that, since this data is unstructured (e.g., raw texts), the semantic
information that the domain experts want to analyze (e.g., emotions in the texts) is not readily available \cite{Zhan2022WhyDY}. Manually annotating these datasets could
cost hundreds to tens of thousands of dollars when using human labor, so social
scientists have turned to machine learning (ML) models \cite{ziems2023large, chenhao2016winning, Zhan2022WhyDY, elsherief2021latent, gabriel-etal-2022-misinfo, baly-etal-2020-detect, ziems-etal-2022-inducing, choi2021meets, demszky2021learning}.
However, applying these ML models is demanding: it involves selecting the correct ML functions and mastering their interfaces, as well as determining appropriate function execution orders when multiple annotations are required \cite{thieme2020machine, grimmer2021machine}.

The second challenge is that these domain experts must turn their questions (often in natural
language) into actual queries, whether using SQL, dataframe libraries, or
statistics libraries. However, this process can be difficult because it involves complex data analytic operations that require advanced programming skills, and the natural language is often underspecified (i.e., \emph{vague}).

To highlight these challenges, we created a new dataset, \datasetname, that
consists of real-world social science research questions, the corresponding
unstructured data, and the ground-truth answers to these questions. \datasetname~ covers all the
topics in the Stanford SALT lab's survey \cite{ziems2023large}, which identifies
core subject areas in social science using ML, as well as the Stanford CS 224C:
NLP for Computational Social Science course \cite{StanfordCS224C}. Among the 120 queries we collected, over half (61) of them require executing two or more ML models, and over a quarter (33) of them are vague.

Existing tools struggle to answer these queries. 
For example,
natural language to SQL (NL2SQL) systems \cite{fu2023catsql, li2024can, copestake1990natural} are unable to handle ML model executions and perform poorly on vague queries with success rates as low as 3\% (Section~\ref{subsec:pass_at_k_exp}), even for NL2SQL systems specifically designed for vague queries at the table schema level \cite{bhaskar2023benchmarking}. 
Generic applications built on large language
models (LLMs) 
also fall short in answering these domain-specific social
science questions, with failure in generating responses in 10 different attempts when querying $Q_{11}$ in \datasetname (Table \ref{tab:dataset_sum}): ``I want to find the emotion triggers for all posts''~\cite{Zhan2022WhyDY} over a CSV file containing 1817 posts in plain text using ChatGPT (GPT-4) \cite{chatgpt, gpt4}.

To address these issues, we propose \sysname, an end-to-end library that assists
social scientists in analyzing unstructured data. 
With the
raw unstructured data and natural language queries as inputs, 
\sysname automatically
parses the natural language, applies the necessary internally supported and user-defined ML functions, and executes the query over the results of the ML models, i.e., structured tables with adequate semantic information.

To accomplish this, \sysname~ uses LLM to parse the natural language query,
decide on the ML functions to apply, and generate the code to perform the
necessary analysis. However, effectively using LLM in \sysname requires overcoming
several challenges.

First, \sysname~ must handle the vague queries that prior NL2SQL systems fail to respond to.
To do this, \sysname~ incorporates a filter (Section~\ref{subsec:filter}) that identifies vague queries, terminates library executions, and suggests specified alternatives to facilitate exploratory processes, which achieves a success rate of 96\% on vague queries.

Second, automatically selecting an appropriate ML function chain is difficult, especially in situations involving complex function dependencies. 
Answering $Q_{11}$ requires first applying the emotion classifier $f_\text{emotion}$ to the posts and then propagating its outputs to the emotion trigger identifier $f_\text{trigger}$.
When directly passing $Q_{11}$ as the user message in the function calling interface provided by
OpenAI using \texttt{gpt-4-0613} \cite{OpenAIsupported} with three functions ($f_\text{emotion}$, $f_\text{trigger}$, and named entity recognizer $f_\text{ner}$) as candidates, functions are called in incorrect orders in 10 different attempts. 

Third, in terms of efficiency, the query cost is high due to the extensive variety of supported ML functions. $Q_{11}$ takes 9,084 input tokens with \sysname's internally supported function list as candidates, exceeding the token limit of 8,192 supported by \texttt{gpt-4-0613} \cite{gpt4}. 

To improve effectiveness, \sysname~ integrates a forward planning mechanism that achieves 98\% accuracy in identifying ML function chains. 
In addition, \sysname~ incorporates doubly linked lists that connect functions with mutual dependencies such as $f_\text{emotion}$ and $f_\text{trigger}$ in $Q_{11}$, which increases the accuracy of queries with implicit function calls from 20\% to 87.7\%.
To improve efficiency, \sysname~ structures the supported function list into a function tree to pass only the functions in a single leaf node as candidates for the function calling interface and inserts alias check blocks that determine whether the annotations to be generated already exist before executing ML functions to prevent redundant executions, saving query costs by 55\%. 
We introduce the details of these components in Section~\ref{subsec:pipeline_selector}. 

We demonstrate the performance and cost efficiency of \sysname~ in Section~\ref{sec:exp}.
\sysname~ achieves 100\% pass @ 3 and 92\% pass @ 1 across all 120 queries in \datasetname. 
The success rate of \sysname~ in responding to vague queries is over 30 times higher than existing NL2SQL systems.
The average cost per query is \$1.06, with only \$0.02 spent on code generation. The end-to-end cost, from vague query reformulation and data annotation to code generation and execution, of \sysname is less than 0.1\% of what traditional social science research spends on data annotation alone.


\section{\datasetname: A Social Science Research Question Dataset} \label{sec:data}
We introduce \datasetname, a dataset containing social science \textbf{q}ueries on \textbf{u}nstructured data \textbf{i}nvoking \textbf{e}xtended \textbf{t}ables with \textbf{ML} models.
\datasetname~ consists of queries that cover all the topics addressed in the Stanford SALT lab's survey \cite{ziems2023large} and the Stanford CS 224C course \cite{StanfordCS224C}. 
\datasetname~ includes 120 queries, spanning 9 prominent social science domains and 25 popular topics, addressing social science problems across 68 sources. 
We present the details in Table ~\ref{tab:dataset_sum}. 

Among these 120 queries, 78 are non-vague queries without unspecified numerical values, 9 are non-vague queries with unspecified numerical values, and 33 are vague queries. 
The 33 vague queries cover the following three common causes of ambiguity in social science research questions:
\begin{enumerate}
\item \textbf{Lack of context}: social scientists pose vague queries with unspecified contexts.
For instance, a psychologist might ask ``Provide cognitive behavioral therapies for these negative thoughts'' ($Q_{16}$), implicitly assuming that ``cognitive behavioral therapies'' refers to positive reframing \cite{CBT}.
\item \textbf{Data insufficiency}: social scientists formulate vague queries that cannot be answered from available data. For instance, ``Is the public mood correlated with, or even predictive of, economic indicators?'' ($Q_2$) \cite{BOLLEN20111} is not answerable by simply analyzing the provided Tweets without incorporating relevant economic statistics.
\item \textbf{Informal or unconventional expressions}: natural language queries may include non-rigid expressions. 
For instance, a social media researcher can ask ``I want to predict whether the conversation will get out of hand'' ($Q_{24}$) \cite{zhang-etal-2018-conversations}, where ``get out of hand'' informally denotes ``become toxic''. 
While the two can be considered equivalent in everyday conversation, the lack of precision in ``get out of hand'' poses challenges for computational analysis.
\end{enumerate}

\datasetname~ provides the unstructured data for each query, consisting of 22,323 data points on average. This data covers a diverse range of unstructured data types, from text and PDF documents to videos. 
On average, each query needs 2.0 semantic annotations for each data point, where 
61 out of the 120 queries require two or more annotations. For evaluation, \datasetname~ includes the ground-truth query results on the provided unstructured data.

\begin{table*}[t!]
    \centering
    \caption{Social Science Domains and Topics Covered in \datasetname}
    \small
    \begin{tabularx}{\textwidth}{cXm{32em}}
    \toprule
    \textbf{Domain} & \textbf{Topic} & \textbf{Example}\\
    \midrule 
    \multirow{3}{*}{Persuasion} & Persuadability \cite{chenhao2016winning} ($Q_1, Q_{41}, Q_{81}, Q_{82}$) & \small Recognize the “malleable” cases. \\
    \cmidrule(lr){2-3}
    & Persuasiveness \cite{Zhan2022WhyDY, yang-etal-2019-lets, Althoff2014HowTA} ($Q_{17}$, $Q_{37}, Q_{47}$, $Q_{77}$)& \small Which posts are persuasive?\\
    \cmidrule(lr){2-3}
    & Attackability \cite{jo-etal-2020-detecting, chenhao2016winning, hidey-etal-2017-analyzing} ($Q_{34}, Q_{74}$)& \small I want to see how sentiment affects the attackability of a sentence.\\
    \midrule
    \multirow{2}{*}{Emotion} & Emotion Classification \cite{BOLLEN20111, saravia-etal-2018-carer, kramer2014exp, hatfield1993emotional, Ferrara1025Measuring} ($Q_2, Q_{42}$) & \small Is the public mood correlated or even predictive of economic indicators? \\
    \cmidrule(lr){2-3}
    & Emotion Triggers \cite{Zhan2022WhyDY} ($Q_{10}$, $Q_{11}, Q_{50}$, $Q_{51}$)& \small I want to find the triggers for all posts of emotion that has the maximum quantity.\\
    \midrule
    \multirow{6}{*}{Social Bias} & Dog Whistles \cite{mendelsohn-etal-2023-dogwhistles} ($Q_3, Q_{43}$) & \small For each persona/in-group, what is the number of each type of dog whistle? \\
    \cmidrule(lr){2-3}
    & Hate Speech \cite{elsherief2021latent, sap2020social, sap-etal-2019-risk, lemmens-etal-2021-improving, mathew2020hate} \newline ($Q_6$, $Q_7$, $Q_8$, $Q_9$, $Q_{46}$, $Q_{47}$, $Q_{48}$, $Q_{49}$)& \small Which posts contain hate speech?\\
    \cmidrule(lr){2-3}
    & Equality \cite{Garg_2018, NIPS2016_a486cd07, prates2019assessinggenderbiasmachine} ($Q_{38}, Q_{78}, Q_{113}$)& \small I want to quantify how the bias of words evolves.\\
    \cmidrule(lr){2-3}
    & Offensiveness \cite{ashida-komachi-2022-towards, chung-etal-2019-conan, fanton-2021-human, chung-etal-2021-knowledge, breitfeller-etal-2019-finding, sap2020socialbiasframes} ($Q_{111}, Q_{112}$)& \small Are the defense to these texts effective?\\
    \midrule
    \multirow{4}{*}{Misinformation} & Fake News \cite{gabriel-etal-2022-misinfo, Lazer2018The} ($Q_4$, $Q_{14}$, $Q_{44}$, $Q_{54}$) & \small Which news headlines contain misinformation? \\
    \cmidrule(lr){2-3}
    & Imaginary Stories \cite{sap-etal-2020-recollection, sap-2022, li2021documentlevel, sprugnoli-tonelli-2019-novel, hurriyetoglu-etal-2021-challenges} \newline ($Q_{25-28}, Q_{65-68}$)& \small I want to get the imaginary stories generated based on the recalled stories. \\
    \cmidrule(lr){2-3}
    & Deceptive Videos \cite{deception-detect, catalina2008separating, Groh_2021} ($Q_{36}, Q_{39}, Q_{76}, Q_{79}$)& \small I want to extract all fake videos. \\
    \midrule
    \multirow{2}{*}{Attitude} & Stance \cite{mohammad-etal-2016-semeval} ($Q_5, Q_{45}$) & \small What is the difference between stance and sentiment? \\
    \cmidrule(lr){2-3}
    & Ideology \cite{baly-etal-2020-detect, iyyer-etal-2014-political, jelveh-etal-2014-detecting, preotiuc-pietro-etal-2017-beyond} ($Q_{29}$, $Q_{30}$, $Q_{69}$, $Q_{70}$)& \small I want to find the percentage of right political ideology. \\
    \midrule
    \multirow{8}{*}{Linguistics} & Figurative Language \cite{chakrabarty-etal-2022-flute, Jacobs2018-JACWMA-2, niculae-danescu-niculescu-mizil-2014-brighter, stowe-etal-2022-impli} \newline ($Q_{12}$, $Q_{13}$, $Q_{52}$, $Q_{53}$) & \small Retrieve the explanations of the premise that entails the figurative sentences. \\
    \cmidrule(lr){2-3}
    & Dialect Features \cite{demszky2021learning, ziems2023multivalue} ($Q_{15}, Q_{55}$)& \small I want to retrieve posts with the most common dialect features. \\
    \cmidrule(lr){2-3}
    & Semantics \cite{pilehvar-camacho-collados-2019-wic} ($Q_{18}, Q_{58}$)& \small I want to retrieve example pairs of the same verb but with different semantics. \\
    \cmidrule(lr){2-3}
    & Discourse Acts \cite{Zhang2017CharacterizingOD, huguet-cabot-etal-2020-pragmatics} ($Q_{19}, Q_{59}$)& \small I want to classify comments in online discussions into a set of coarse discourse acts toward the goal of better understanding discussions at scale. \\
    \cmidrule(lr){2-3}
    & Echo Chamber Effect \cite{Alatawi2023Quantifying} ($Q_{114}, Q_{115}$)& \small Extract tweets with low Echo Chamber Effect. \\
    \midrule
    \multirow{4}{*}{Psychology} & Mental Health \cite{ziems-etal-2022-inducing, rothbaum2000cognitive, asai2018happydb, mcrae2020emotion, sharma2020computational, liu2021emotional} \newline ($Q_{16}$, $Q_{20}$, $Q_{21}$, $Q_{32}$, $Q_{33}$, $Q_{56}$, $Q_{60}$, $Q_{61}$, $Q_{72}$, $Q_{73}$) & \small Provide cognitive-behavioral therapies for these negative thoughts. \\
    \cmidrule(lr){2-3}
    & Social Psychology \cite{Danescu-Niculescu-Mizil+al:13b, weller-seppi-2019-humor, Li2020Studying, brown1987politeness, zhang-etal-2018-conversations} \newline ($Q_{22}, Q_{23}$, $Q_{24}$, $Q_{62}, Q_{63}$, $Q_{64}$)& \small I want to identify all impolite posts. \\
    \midrule
    \multirow{2}{*}{Social Roles} & Trope \cite{bamman-etal-2013-learning, chu2018learning} ($Q_{31}, Q_{71}$) & \small I want to find all characters that are chanteuse. \\
    \cmidrule(lr){2-3}
    & Relationship \cite{choi2021meets, Danescu-Niculescu-Mizil+al:12a} ($Q_{35}, Q_{40}, Q_{75}, Q_{80}$)& \small I want to study how conversational behavior can reveal power relationships. \\
    \midrule
    \multirow{2}{*}{Legal Services} & Documents\cite{caselaw} ($Q_{90}, Q_{91}, Q_{94-97}, Q_{102-110}, Q_{116-120}$) & \small I want to summarize the documents. \\
    \cmidrule(lr){2-3}
    & Texts \cite{Engstrom2024} ($Q_{83-89}, Q_{92}, Q_{93}, Q_{98-101}$)& \small I want to give an overview of the cases. \\
    \bottomrule
    \end{tabularx}
    \label{tab:dataset_sum}
\end{table*}
\section{Use Cases}\label{sec:prob}


We demonstrate how an end-to-end library helps social scientists handle the three types of queries in \datasetname.

\minihead{Non-vague queries without unspecified numerical values} A media researcher collects Tweets with dog whistles and asks ``For each (targeted) persona/in-group, I want to know the number of each type of dog whistles.'' ($Q_3$) ~\cite{mendelsohn-etal-2023-dogwhistles} 

The library first loads the Tweets as a 
single-column table, then proceeds to select and apply ML functions to extend the table with necessary semantic information to answer the query. In this case, $f_\text{p/i}$ that identifies the targeted persona/in-group and $f_\text{dw-type}$ that classifies dog whistle types should be executed as explicitly stated in the query. However, $f_\text{dw}$ that extracts the dog whistle terms should first be applied to the Tweets, since the output of $f_\text{dw}$ implicitly serves as the input of $f_\text{p/i}$ and $f_\text{dw-type}$. The detailed dependencies can be viewed in Figure ~\ref{fig:ui}. 

When the table is extended with adequate semantic information, the query is translated into code similar to the following SQL code.
\begin{lstlisting}[language=CustomSQL]
SELECT persona_or_ingroup, type, COUNT(*) AS count
FROM table
GROUP BY persona_or_ingroup, type
\end{lstlisting}

Once the extended table and code are generated, the library executes the code and displays the execution results.

\minihead{Non-vague queries with unspecified numerical Values}\label{subsec:non-vague2} A sociolinguist collects Reddit posts where people argue and aims to filter the posts according to persuasion effect scores by asking ``Which posts are persuasive?'' ($Q_{17}$) ~\cite{chenhao2016winning}

To decide whether a post is considered ``persuasive'', the user should specify a persuasion effect score as the criterion. However, the user may be unsure about this value before knowing the score distributions. Therefore, the library issues a warning and lets the user decide whether to proceed with this value remaining unspecified or to input a new query with specified numerical values such as ``Which posts have a persuasion effect score > 0.9?''

Regardless of which query the user chooses to proceed with, the library first loads the posts as a single-column table and appends an additional column containing persuasion effect scores from applying the persuasion effect score calculator $f_\text{pe}$ on the posts.
For code generation, the library generates code similar to the following SQL code.
\begin{lstlisting}[language=CustomSQL]
SELECT * FROM table 
WHERE persuasion_effect_score > @X@
\end{lstlisting}
where \textcolor{red}{\texttt{X}} is either the user-specified value, if provided, or the library chooses a reasonable value based on the data distribution. Finally, the library executes the code and displays the results.



\minihead{Vague queries} An economist collects Tweets to extract emotions and analyze their correlation with economic conditions by posing an exploratory query: ``Is the public mood correlated with, or even predictive of, economic indicators?'' ($Q_2$) \cite{BOLLEN20111} 

The library detects the vagueness, terminates the execution, and suggests alternative queries that (1) are specific enough and (2) align with user intent. For example, an alternative query could be ``I want to compute the emotion distribution of the posts.''

With the recommended non-vague query, the library first loads the Tweets as a single-column table and appends an additional column containing emotion classes from applying the emotion classifier $f_\text{emotion}$ on the posts. The library then generates code similar to the following SQL code.
\begin{lstlisting}[language=CustomSQL]
SELECT emotion, COUNT(emotion) AS count 
FROM table 
GROUP BY emotion
\end{lstlisting}

Finally, the library executes the code and displays the results.

\section{\sysname: An End-to-end Library For Social Science Research Questions}\label{sec:method}
\begin{figure}[t!]
    \graphicspath{{figures/}}
    \centering
    \includegraphics[width=0.45\textwidth]{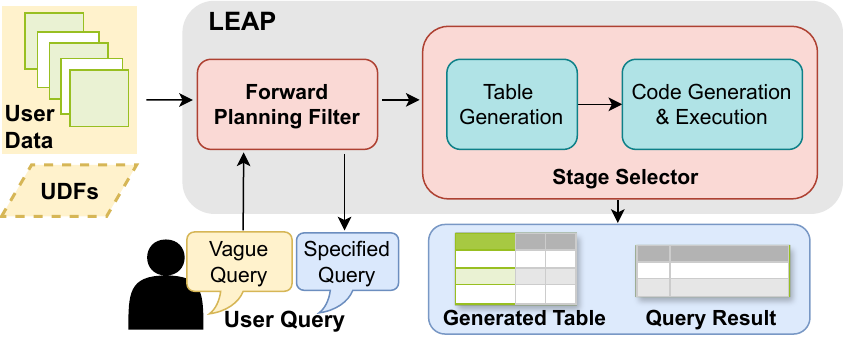}
    \caption{Workflow of \sysname.}
    \label{fig:overall}%
\end{figure}

In this section, we propose \sysname, an \textbf{L}LM-powered \textbf{e}nd-to-end \textbf{a}utomatic library for \textbf{p}rocessing social science research questions. We provide an overview of the workflow of \sysname~ in Figure ~\ref{fig:overall}. 

\sysname~ takes user-provided data $\mathcal{D}$, a potentially vague query in natural language $q$ on $\mathcal{D}$,
and optionally, the user-defined functions (UDFs), as input. It processes these inputs to generate result $r$ in response to $q$ together with $\mathcal{T}$, a structured table that contains adequate semantic information.
$\mathcal{D}$ can either be unstructured data like the raw data in \datasetname, which \sysname~ loads as a single-column table, or it can readily be a structured table containing partial or full semantic information.

\sysname~ consists of (1) a forward planning filter (Section~\ref{subsec:filter}) that determines if $q$ is vague, and (2) a stage selector (Section~\ref{subsec:pipeline_selector}) that selects among candidate stages of \emph{table generation} (Section~\ref{stage:table_gen}), \emph{code generation} (Section~\ref{stage:code_gen}), \emph{code execution} (Section~\ref{stage:code_exec_result_eval}), and \emph{result display} (Section~\ref{stage:code_exec_result_eval}). We introduce the user interface of \sysname, including the integration of UDFs, in Section~\ref{subsec:user_interface}.

\subsection{Forward Planning Filter}\label{subsec:filter}

To prevent users from multiple aimless retries when giving vague queries, \sysname~ starts with a forward planning filter $FP$. Given the full function list $\mathcal{F}$, 
which consists of \sysname's internally supported functions and any UDFs, 
and user-provided data $\mathcal{D}$,
$FP$ decides whether the user query $q$ is vague through a query check. 

$FP$ prompts \texttt{gpt-4-0613} \cite{OpenAIsupported} and checks
(i) whether there exists a well-defined function chain to annotate the data (i.e., whether $\mathcal{D}$ can be extended into deterministic structured tables), (ii) whether $q$ can be answered through a deterministic sequence of SQL operations given a table with adequate information, and (iii) whether there are any unspecified numerical values (i.e., whether $q$ can be translated into executable code). 
$FP(q, \mathcal{F}, \mathcal{D})$ falls into three cases:
\begin{enumerate}
    \item When $q$ is clear in terms of (i), (ii), and (iii), $FP$ passes the query check and generates a planned function chain $\mathcal{C}$. 
    \item When $q$ satisfies only (i) and (ii), $FP$ generates a planned function chain $\mathcal{C}$ and a warning message where users can decide whether to proceed or not. If users decide to proceed, $FP$ passes the query check and lets the \emph{code generation} (Section~\ref{stage:code_gen}) stage handle the unspecified numerical values.
    \item When $q$ does not meet the criteria of either (i) or (ii), $FP$ identifies $q$ as a vague query and fails the query check by terminating the execution and generating an alternative query list $\mathcal{Q}$ based on $\mathcal{F}$, $\mathcal{D}$, and $q$.
\end{enumerate}

In cases (1) and (2), $FP$ falls back to returning an empty function chain $\mathcal{C} = \varnothing$ when $\mathcal{D}$ already contains adequate semantic information, such that $q$ can be directly translated into SQL code. For example, if $q$ extracts Tweets containing a specific \texttt{keyword}, it can be answered with code similar to 
\begin{lstlisting}[language=CustomSQL]
SELECT * FROM table WHERE Tweet LIKE '%keyword%'
\end{lstlisting} 
without applying ML functions to \texttt{Tweet}.

We use chain of thoughts (COT) \cite{wei2023chainofthought}, a forward planning prompt technique that decomposes large tasks into small steps, to help $FP$ plan for the function chain $\mathcal{C}$, as well as few-shot learning \cite{brown2020language}, a prompt technique that provides a limited number of examples to guide the LLMs, to improve query recommendations.


\subsection{Stage Selector}\label{subsec:pipeline_selector}
The user query $q$ that passes $FP$ and enters the stage selector is non-vague. 
The stage selector follows the typical steps of data analysis in social science: first, annotate the raw data with the necessary semantic information; then, write code based on the research question; when the code is ready, execute it in the correct setup; once the result is ready, display it for social scientists to view and evaluate.
The stage selector maintains a progress record $R$ to track the executed stages. The stage selector automatically selects the next stage to be executed according to $q$, the current table, and $R$. The stage selector enters the \emph{table generation} stage (Section~\ref{stage:table_gen}) when the current table is incomplete, and enters the \emph{code generation} stage (Section~\ref{stage:code_gen}) otherwise. If given a complete table and executable code, \sysname proceeds to the \emph{code execution} stage (Section~\ref{stage:code_exec_result_eval}). Upon detecting an execution result, \sysname moves to the final \emph{result display} stage (Section~\ref{stage:code_exec_result_eval}). \sysname~ automatically terminates after entering the same stage three times consecutively, providing detailed feedback for users to refine $q$. 
We use the function calling interface provided by OpenAI API with \texttt{gpt-4-0613} \cite{OpenAIsupported}, where each stage is wrapped as an individual function call.
\subsubsection{Table Generation}\label{stage:table_gen}
 \begin{figure}[t!]
    \graphicspath{{figures/}}
    \centering
    \includegraphics[width=0.37\textwidth]{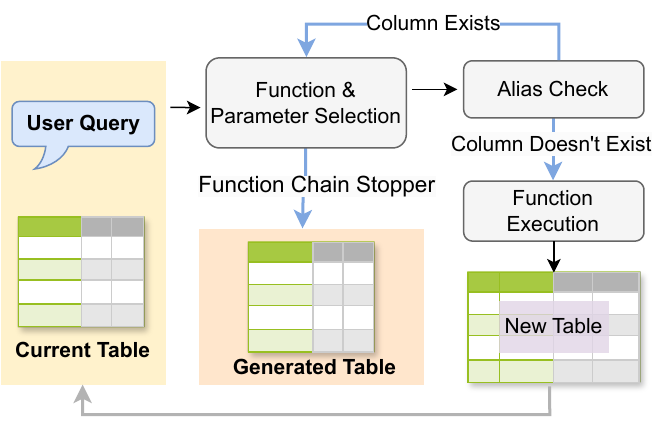}
    \caption{Workflow of the table generation stage.}
    \label{fig:table_gen}%
\end{figure}
In this stage, \sysname generates a structured table $\mathcal{T}$ and corresponding column descriptions by extending the current table with columns annotated with ML functions, ensuring adequate and correct information is included to answer $q$ (Figure ~\ref{fig:table_gen}). 

We use the function calling interface provided by OpenAI API with \texttt{gpt-4-0613} \cite{OpenAIsupported}. 
Inspired by the idea of COT \cite{wei2023chainofthought}, we deploy a step-to-step example as guidance. Specifically, we integrate an example query ``I want to count the number of positive paragraphs in the PDF document.'' as part of the prompt. For each progressive status indicated by existing column descriptions, the prompt includes the ground-truth function selections in order ($f_\text{ocr}$: OCR function that translates from pdf files to texts; $f_\text{para}$: paragraph separator; $f_\text{sentiment}$: sentiment analyzer applied on each paragraph; $f_\text{stopper}$: function call chain stopper), together with the ground-truth parameter selections (i.e., the columns to apply these functions). 

\begin{figure}[t!]
    \graphicspath{{figures/}}
    \centering
    \includegraphics[width=0.36\textwidth]{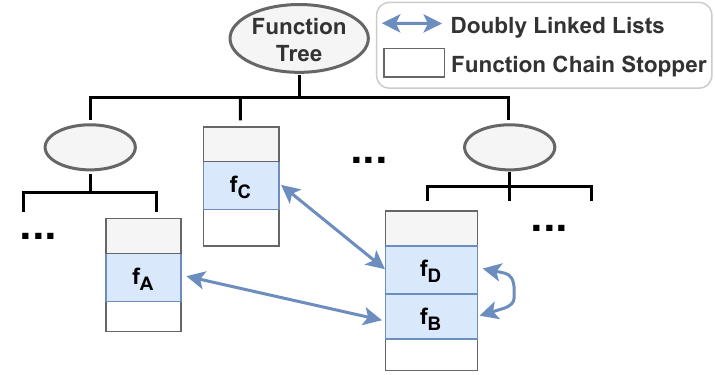}
    \caption{Structure of the supported function list $\mathcal{F}$ in \sysname.}
    \label{fig:func-struct}%
\end{figure}

To further optimize function call accuracy and minimize query costs, the function list $\mathcal{F}$ is organized into a \emph{function tree}, where mutually dependent functions form \emph{doubly linked lists} (Figure \ref{fig:func-struct}). 

\minihead{Function tree}
To support a wide range of user queries, the function list size $|\mathcal{F}|$ is enormous such that it exceeds the token limit of 8,192 for \texttt{gpt-4-0613} even without any UDFs when the entire list becomes candidates.
To overcome the issue, we organize $\mathcal{F}$ into a function tree where functions are grouped into subgroups based on their types and stored in leaf nodes (Figure \ref{fig:func-struct}).
Every leaf node includes a stopper function $f_\text{stopper}$ for the function chain to end when the current table includes adequate information. \sysname goes through a tree search process based on the hierarchical structure, and only the functions in the targeted leaf node become the candidates for the function calling interface.

\minihead{Doubly linked lists}
The ML function executions can be mutually dependent, but user queries are often implicit with such dependencies. For example, a social media researcher asks ``I want to get the readers' actions of headlines that have a high rate of likelihood to spread'' ($Q_{14}$) \cite{Gabriel2022MisinfoRF}. To get readers' actions using $f_\text{reader-action}$, the writers' intent should first be identified using $f_\text{writer-intent}$, and to assess a headline's likelihood of spreading using $f_\text{spread-likelihood}$, the reader perception should first be inferred using $f_\text{reader-perception}$. 

To resolve the issue, we connect dependent functions with doubly linked lists (Figure \ref{fig:func-struct}), 
where a forward pointer $f_A \rightarrow f_B$ represents that the outputs of $f_A$ is necessary for the execution of $f_B$, and a backward pointer $f_C \leftarrow f_D$ represents that the execution of $f_D$ depends on the outputs of $f_C$. 
This design reveals the underlying relationships among functions, and connects different leaf nodes based on function dependencies, resulting in a more accurate tree path search. 
The doubly linked lists allow users to obtain the correct results without having to know or explicitly state the implicit function dependencies, which are easy to neglect.

\minihead{Alias check blocks} 
ML functions can be repeatedly called, which is especially common in cases where (1) complex function dependencies exist, and/or (2) $\mathcal{D}$ contains information derived from ML functions. For example, a literature scholar can ask ``I want to get the imaginary stories generated based on the recalled stories.'' \cite{sap-etal-2020-recollection}. 
To do so, one first summarizes the provided recalled stories using the summarizer $f_1$, and then generates imaginary stories based on the summaries using the story generator $f_2$. Given that summarizing texts is easier and cheaper than generating them, users can readily provide the summaries of the recalled stories, i.e., the outputs of $f_1$, in $\mathcal{D}$. However, due to the function dependencies indicated by $f_1 \leftarrow f_2$, where the execution of $f_2$ depends on the outputs of $f_1$, executing $f_2$ can result in repeated calls to $f_1$. 
Although the correctness of the final results is not affected by such repetitive execution, it is inefficient both in resources, since ML functions are expensive to execute, and in time, since users have to wait for additional rounds of ML function execution time. 


To avoid this, \sysname~ inserts an alias check block before the execution of each function to determine if any column $c_\text{current}$ already contains information that exactly matches the outputs to be generated. In such cases, columns are aliased as $c_\text{new} = c_\text{current}$ instead of generating new ones. This also generates more organized tables $\mathcal{T}$, preventing confusion in subsequent stages.

\subsubsection{Code Generation}\label{stage:code_gen}
In this stage, \sysname~ translates the users' natural language query $q$ into executable code on the extended table $\mathcal{T}$. The code generator generates code that assigns the final execution result to a user-defined variable. 
The code generator is an NL2SQL system using \texttt{gpt-4-0613} \cite{OpenAIsupported}. It takes the column descriptions and, if any, sample values of $\mathcal{T}$ along with $q$ as input, and produces executable code that answers $q$ on $\mathcal{T}$ as output.
If $q$ contains unspecified numerical values, the code generator automatically chooses a reasonable value based on the data distribution.


\subsubsection{Code Execution and Result Display}\label{stage:code_exec_result_eval}
With the extended table $\mathcal{T}$ and corresponding code prepared to address user queries, \sysname~ directly executes the code and displays the execution results.
The code executor executes the generated code on the extended table $\mathcal{T}$. The result display function reads and displays the result.

\subsection{User Interface}\label{subsec:user_interface}
After installing \sysname via commands like \texttt{pip install}, which are typically familiar to users, they can initiate and execute the entire process with a single function signature:
\begin{lstlisting}[style=CustomPython, breaklines=false]
result, table = leap(query, data, description),
\end{lstlisting}
where \texttt{result} is the response to the user query in natural language \texttt{query} based on the user-provided data \texttt{data}, and \texttt{table} is the table containing adequate semantic information annotated by ML functions. \texttt{description} contains descriptions of \texttt{data}.

\begin{figure}[t!]
    \centering
    \includegraphics[width=0.48\textwidth]{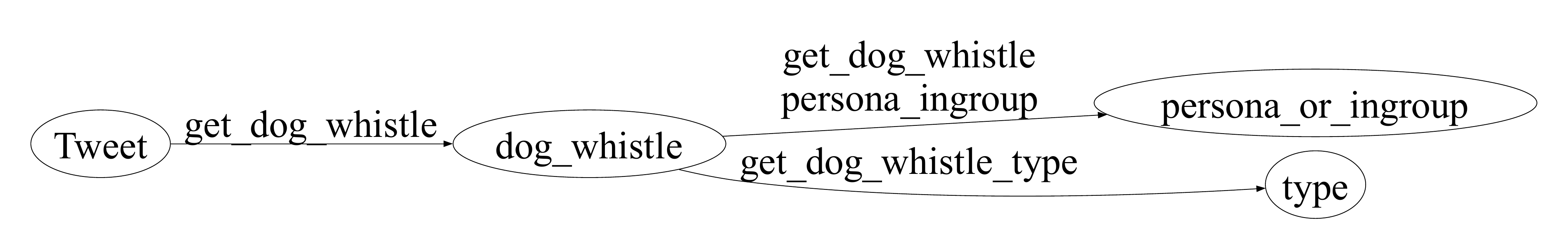}
    \caption{\sysname~ table generation stage's display for $Q_3$: ``For each (targeted) persona/in-group, I want to know the number of each type of dog whistle.''~\cite{mendelsohn-etal-2023-dogwhistles}}
    \label{fig:ui}
\end{figure}

\sysname~ displays the current stage during its execution to indicate the progress.  
During the table generation stage, \sysname~ dynamically updates the column mapping relation graph (Figure \ref{fig:ui}).


To include UDFs to answer a query, users simply need to pass a list containing function metadata, \texttt{udf\_metadata}, as a parameter in the \texttt{leap} function.
\sysname~ integrates UDFs with the internally supported function list and executes them in user space.

We invited a legal researcher to use \sysname~ for their research. They found that \sysname~ streamlines legal document annotations and enhances time efficiency for vague queries. They also found \sysname~ easy to install, load, and use with an intuitive interface.
\section{Experiments}\label{sec:exp}
In this section, we evaluate the applicability of \sysname~ and its components. 
We introduce the experiment setup in Section~\ref{subsec:exp_setup}. We demonstrate the applicability of \sysname~ in social science research in Section~\ref{subsec:pass_at_k_exp}. We explore the effectiveness of the critical components in Section~\ref{subsec:component_studies} and conduct ablation studies in Section~\ref{subsec:ablation_studies}. 

\subsection{Experiment Setup} \label{subsec:exp_setup}
We outline the metrics and baselines for evaluation.

\minihead{Metrics} We use \emph{pass @ k} \cite{chen2021evaluating, kulal2019spoc}, where a query is considered correctly answered if the result from any of the $k$ executions matches the ground truth, as our primary performance metric. We select $k = 1, 3, 5$. With the goal of improving efficiency in social science research, we place a greater emphasis on $k = 1$.
We also track the average API cost for all requests in answering each query \cite{openai-price}.

\minihead{Baselines} 
To demonstrate the superiority of \sysname~ compared to existing NL2SQL and question answering systems, we select $9$ representative baselines.
We select \textbf{(1)} TAPAS large model fine-tuned on WikiTable Questions \cite{tapas-huggingface, herzig2020tapas, eisenschlos2020understanding, DBLP:journals/corr/PasupatL15} (TAPAS), the state of the art for table question answering tasks. WikiSQL \cite{zhongSeq2SQL2017} and Spider \cite{yu-etal-2018-spider} are the two major datasets for training and evaluating NL2SQL systems. We select \textbf{(2)} TAPEX \cite{liu2022tapextablepretraininglearning}, the state of the art on WikiSQL, \textbf{(3)} T5 finetuned on WikiSQL (T5-WikiSQL) \cite{wikisql-huggingface, raffel2023exploring}, the most downloaded model on Huggingface as of 02/25/2024 finetuned on WikiSQL for NL2SQL tasks, \textbf{(4)} DAIL-SQL + GPT-4 + Self-Consistency (DAIL-SQL) \cite{dail_sql}, the state of the art on Spider, and \textbf{(5)} T5 finetuned on Spider (T5-Spider) \cite{spider-huggingface, raffel2023exploring}, the most downloaded model on Huggingface as of 02/25/2024 finetuned on Spider for NL2SQL tasks. To compare the performance of \sysname~ with existing NL2SQL systems designed for vague queries, we select \textbf{(6)} LogicalBeam \cite{bhaskar2023benchmarking}, which successfully handles vague queries at the table schema level.
To show the effectiveness of the design of \sysname, we select \textbf{(7)} \texttt{gpt-4-0613} (GPT-4), the LLM prompted by \sysname's code generator, including a performance comparison with \textbf{(8)} \texttt{gpt-3.5-turbo-0125} (GPT-3.5), both prompted with the official NL2SQL prompt \cite{openai-nl2sql-prompt}.
To compare \sysname with question answering systems on unstructured data, we select \textbf{(9)} RoberTa + Parallel + Adapters (Roberta) \cite{roberta, roberta-huggingface}, the latest model on SQuAD2.0 \cite{rajpurkar2018knowdontknowunanswerable}, a benchmark dataset consisting of question and answer pairs, including unanswerable questions,  based on Wikipedia articles.

\subsection{\sysname~ Achieves Reliable Performance With Low Cost} 
\label{subsec:pass_at_k_exp}
We demonstrate the performance and cost efficiency of \sysname. 

\minihead{\sysname~ achieves a 92\% accuracy}
We first examine the performance of \sysname~ in various social science domains.
We run \sysname $5$ times over each of the $120$ queries in \datasetname~ to obtain pass @ 3, pass @ 5, and the average pass @ 1. 
For vague queries, a run is successful if \sysname~ correctly rejects the query and recommends alternatives, one of which yields a result that matches the ground truth.
For non-vague queries with unspecified numerical values, a run is successful if \sysname~ generates a warning, and the result of the query with specified numerical values matches the ground truth.

\begin{figure}[t!]
    \graphicspath{{figures/}}
    \centering
    \includegraphics[width=\columnwidth]{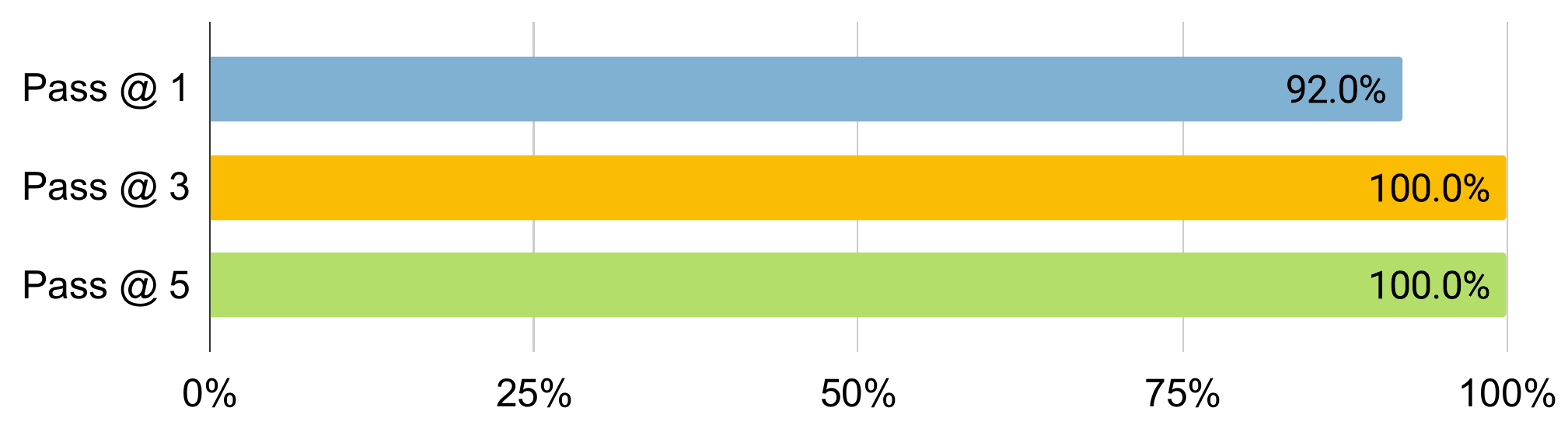}
    \caption{Pass @ k of \sysname over \datasetname.}
    \label{fig:performance}%
\end{figure}

As we show in Figure \ref{fig:performance}, \sysname~ achieves an average of 92\% pass @ 1, where 79 out of 120 queries achieve a 5 out of 5 success rate, 34 achieve a 4 out of 5 success rate, and the remaining 7 achieve a 3 out of 5 success rate. 
The average pass @ 1 achieves 91.5\% for non-vague queries, and 93.3\% for vague queries. 
\sysname~ achieves 100\% for both pass @ 3 and pass @ 5.
\sysname consistently performs well despite the large variance in query complexity across different social science research questions.

\minihead{\sysname~ outperforms baselines in responding to vague queries} \label{subsec:baseline}
We compare the performance of \sysname~ to $7$ NL2SQL systems and $2$ question answering systems on vague queries. 
For \sysname and baselines (1)-(8), we run the $33$ vague queries in \datasetname~ on the pre-generated structured tables with adequate information. Roberta is provided with texts following its original configuration.
We perform $5$ runs for each query. 

\begin{figure}[t!]
    \graphicspath{{figures/}}
    \centering
    \includegraphics[width=0.49\textwidth]{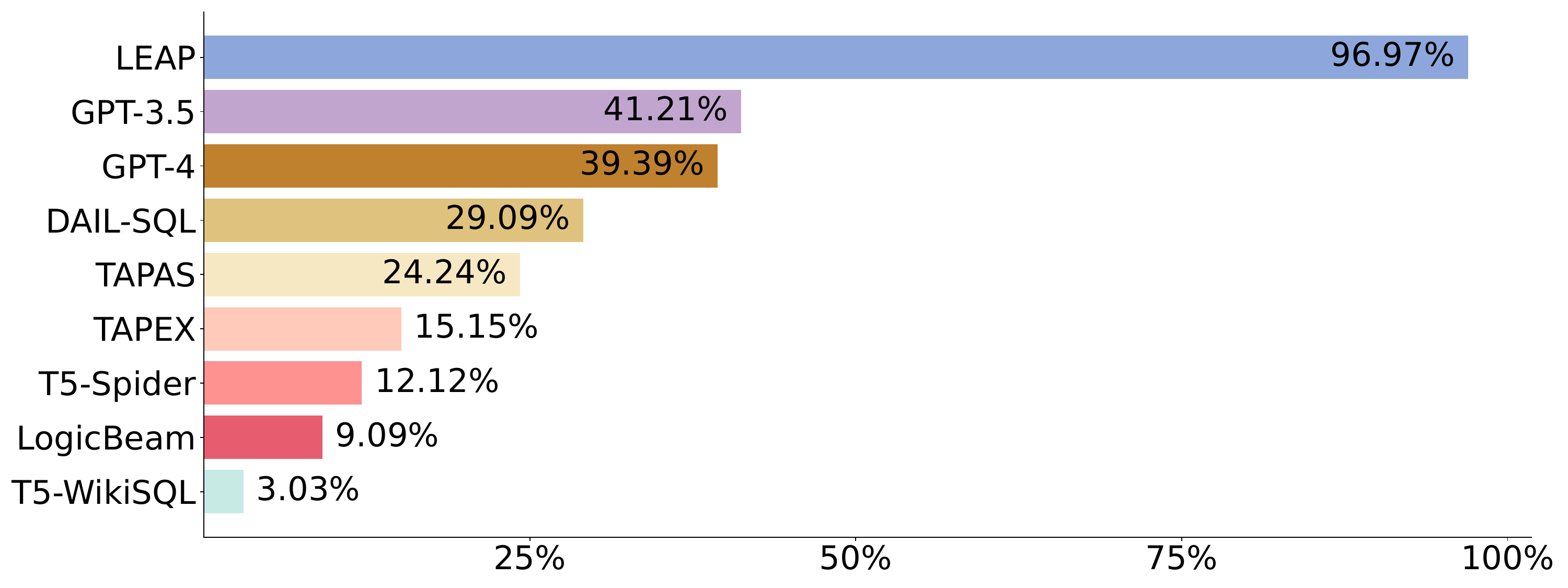}
    \caption{Pass @ 1 of \sysname and baselines over the 33 vague queries in \datasetname on structured tables.
    }
    \label{fig:baseline}%
\end{figure}

As we show in Figure \ref{fig:baseline}, \sysname~ succeeds in producing correct results in 96.97\% of all runs, GPT-3.5 succeeds in 41.21\% of all runs, GPT-4 succeeds in 39.39\% of all runs, DAIL-SQL succeeds in 29.09\% of all runs, TAPAS succeeds in 24.24\% of all runs, and TAPEX, T5-Spider, LogicalBeam, and T5-WikiSQL have accuracy of 15.15\%, 12.12\%, 9.09\%, and 3.03\% respectively. Roberta has an accuracy of 3.03\%, while \sysname achieves an accuracy of 93.3\% on the 33 vague queries when also provided with unstructured data.


\sysname outperforms existing systems due to its capability to identify common vagueness in social science queries and generate alternative specified queries.
The execution results of LogicalBeam and Roberta indicate that the ambiguity in social science queries extends beyond the table schema level and a limited number of text segments, where \sysname~ demonstrates a significant advantage.

\begin{figure}[t!]
    \graphicspath{{figures/}}
    \centering
    \includegraphics[width=\columnwidth]{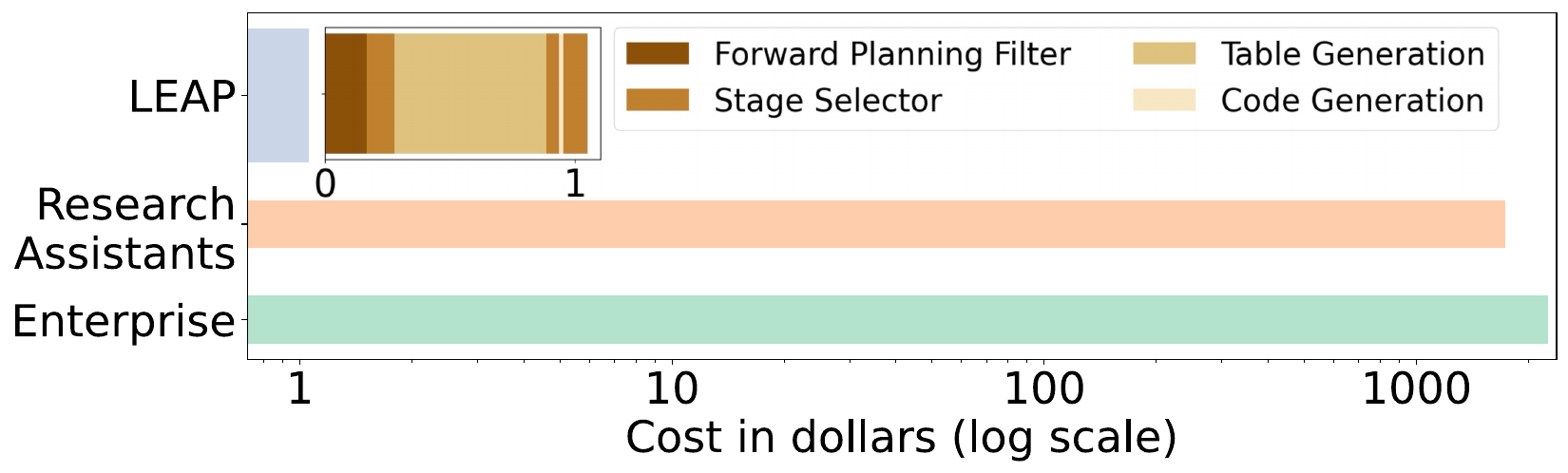}
    \caption{Cost breakdown of \sysname~ and comparison with traditional social science research methods.}
    \label{fig:cost}%
\end{figure}

\minihead{\sysname~ is cost-efficient} 
We compare the query cost of \sysname~ with traditional social science research methods. 
The end-to-end execution cost of \sysname on \datasetname queries is \$1.06 per query on average. Specifically, the cost of the forward planning filter is \$0.17, the stage selector is \$0.22, the table generation stage is \$0.66, and the code generation stage is \$0.02 (Figure ~\ref{fig:cost}).

When estimating the costs of traditional social science research methods, we consider only the data annotation costs, excluding all other costs such as labor and time for code writing and execution, and assuming a specified research question is readily formulated.
We estimate traditional data annotation costs based on the average size of the unstructured data with 22,323 data points, and on average each query requires $2.0$ new annotations for each data point. We calculate the cost in two distinct ways: (1) with professional label providers from enterprises like Scale \cite{scale}, each annotation costs \$0.05, yielding a total cost of \$2,265.78 per query; and (2) by hiring research assistants (RAs) at the minimum wage in Illinois of \$14 per hour. Assuming it takes 10 seconds for an RA to provide an annotation, which is conservative given the extensive volume of videos and documents, it costs \$1,736.23 per query. 

As we show in Figure ~\ref{fig:cost}, it costs less than 1/1000 to answer a query using \sysname~ than using traditional social science research methods. 
This low cost results from \sysname's efficient prompt designs with various structures, 
as we demonstrate in Sections \ref{subsec:component_studies} and \ref{subsec:ablation_studies}.

\subsection{Component Studies} \label{subsec:component_studies}
We study the effectiveness of the critical components in \sysname.

\minihead{Forward planning filter achieves over 96\% accuracy}
We study the accuracy of the results of the forward planning filter. We first define which results are considered accurate case by case.
\begin{enumerate}
    \item For non-vague queries without unspecified numerical values, the forward planning filter should not signal unnecessary warnings or incorrectly label these queries as vague, and should accurately identify function chains.
    \item For non-vague queries with unspecified numerical values, the forward planning filter should issue warnings inquiring whether the user intends to proceed, and should accurately identify function chains.
    \item For vague queries, the forward planning filter should detect the vagueness and terminate the execution. The forward planning filter should recommend alternative queries, at least one of which satisfies condition (1) or (2).
\end{enumerate}

We run \sysname~ 5 times for each query, and the accuracy for all 120 queries achieves 96.5\%. The accuracy for the non-vague queries without unspecified numerical values achieves 96.9\%, the accuracy for the non-vague queries with unspecified numerical values achieves 97.8\%, and the accuracy for the vague queries achieves 95.8\%. The accuracy for vague queries is lower due to the uncertainty introduced by generating specified queries, including those that do not match the user's original intentions.

\minihead{Stage selector achieves 99\% accuracy} We study the accuracy of the stage selector. A run is accurate if all necessary stages are selected and executed without errors or repetitions. We run \sysname 5 times over all 120 queries. The accuracy of the stage selector achieves 98.5\%. 
Out of the 9 failure cases, only 1 is due to redundant selection, and the remaining 8 are due to inaccurate selection or erroneous execution. 
The stable performance of the stage selector ensures the workflow of \sysname is logically and efficiently divided.

\minihead{Function selection achieves 98\% accuracy}
To evaluate the capability of \sysname~ in calling appropriate functions, we study the accuracy of function selection.
The function selection of a query is considered accurate if (1) the function tree search engine identifies correct paths to the tree leaves, (2) the correct functions are called in appropriate orders, and (3) the function chain stopper $f_\text{stopper}$ is correctly called. 
We run \sysname~ 5 times over all 120 queries. The accuracy of function selection achieves 97.8\%.
The prompt designs for function calling in \sysname's table generation stage ensure correct data annotations in social science research.

\minihead{Parameter selection achieves 99\% accuracy}
We further explore \sysname's capability in mapping relationships among columns by analyzing the accuracy of parameter selection.
The parameter selection is considered accurate if the correct columns are selected to derive the new columns.
We run \sysname~ 5 times for all 120 queries. The accuracy of parameter selection achieves 98.8\%.
The table column descriptions are properly maintained during \sysname's table generation stage to ensure correct column mappings.

\minihead{Alias check blocks work with forward planning filter to ensure 0\% redundancy}
We study the capability of \sysname~ to minimize unnecessary ML function executions. We select the 20 queries in \datasetname~ that require multiple function calls with mutual dependencies, and extend the data with intermediate annotations. For tests in Section~\ref{subsec:pass_at_k_exp} that provide unstructured data \texttt{X}, and where the ML functions annotate this data into a table with columns \texttt{[X, $f_1$(X), $f_2$($f_1$(X))]}, we provide a structured table with columns \texttt{[X, Y]}, where \texttt{Y = $f_1$(X)}. We examine whether the ML functions that generate these intermediate annotations (i.e., the \texttt{$f_1$}'s) are redundantly executed.
We run each of these 20 queries 5 times. 

Of the 94 runs (accuracy 94\%) that produces correct results, all (100\%) successfully avoid redundant ML function executions. The forward planning filter avoids the calling of these functions by excluding them from the planned function chains in 54.3\% of the runs, and alias check blocks avoid the execution of these functions in all the remaining 45.7\% when they are redundantly called. 
In 11 out of 20 queries, alias check blocks avoid the execution of unnecessary ML functions in at least 3 out of the 5 total runs. \sysname~ avoids all (100\%) of the unnecessary function executions when given multiple intermediate columns. By accurately identifying redundancy in annotations, \sysname minimizes ML function execution costs.


\subsection{Ablation Studies} \label{subsec:ablation_studies}
We conduct ablation studies to reveal the impact of each component. 

\minihead{Forward planning filter doubles accuracy}
We study how the forward planning filter affects \sysname's performance by removing it and running the same tests as Section~\ref{subsec:pass_at_k_exp} on all 120 queries. 

With the forward planning filter removed, the average pass @ 1 drops from 92\% to 43.7\%, where only 19 out of 120 queries achieve a 5 out of 5 success rate, which is less than a quarter of the number of \sysname~ with the forward planning filter. 
Specifically, the average pass @ 1 for non-vague queries drops from 91.5\% to 54.7\%, and the average pass @ 1 for vague queries drops from 93.3\% to 14.5\%. 
The pass @ 3 drops from 100\% to 59.2\%, where pass @ 3 for non-vague queries drops to 73.6\%, and pass @ 3 for vague queries drops to 21.2\%.
The pass @ 5 drops from 100\% to 66.7\%, where pass @ 5 for non-vague queries drops to 78.2\%, and pass @ 5 for vague queries drops to 36.4\%. 
This indicates that the forward planning filter contributes both to specifying vague queries and identifying complex function chains, with the former having greater influence.

\minihead{Doubly linked lists raise accuracy by 4$\times$}
We examine how doubly linked lists help answer queries that require executing mutually dependent functions. We remove all doubly linked lists and run the same tests as Section~\ref{subsec:pass_at_k_exp} on all 13 queries whose function chains involve functions in these doubly linked lists. 

Doubly linked lists provide explicit information about function dependencies, which otherwise can only be inferred from column descriptions. With all doubly linked lists removed, the average pass @ 1 drops from 87.7\% to 20\%, the pass @ 3 drops from 100\% to 46.2\%, and the pass @ 5 drops from 100\% to 61.5\%.



\minihead{Function tree halves query costs}
We study how the function tree structure lowers query cost. 
We remove the function tree search engine and pass all internally supported ML functions as candidates for the function calling interface. 
We measure query costs in terms of the total number of prompt tokens used in table generation, the stage where the function tree structure contributes. 

Instead of the entire function list, \sysname only selects from a small subset of functions (i.e., the selected tree leaf nodes) for each function call. This reduction in prompt tokens outweighs the additional prompts used by the tree search engine, especially when multiple function calls (i.e., multiple annotations) are required. By removing the function tree structure, the average number of prompt tokens increases from 20,861 to 46,318, resulting in the total query cost being 122.03\% higher than that of the original \sysname~ with the function tree structure in the table generation stage. 



\minihead{Function tree raises accuracy by 5$\times$}
We study how the function tree structure affects \sysname's performance by modifying \sysname~ as in the previous section and running the same tests as Section~\ref{subsec:pass_at_k_exp} 
on all 120 queries. 

The function tree's removal impedes the selection of the function chain stopper $f_\text{stopper}$ among the extensive pool of candidates. The average pass @ 1 drops from 92\% to 16.2\%, pass @ 3 drops from 100\% to 34.2\%, and pass @ 5 drops from 100\% to 39.2\%. 
\section{Related Work}\label{sec:related}
We review related work from the following two aspects.

\minihead{Computational social science} 
As available data scales, social scientists increasingly quantify complex social science problems and leverage computational resources for solutions \cite{lazer2020computational}. This leads to the development of the computational social science field \cite{lazer2009computational, annurev-soc-121919-054621}, 
enabling social scientists to derive deep insights from large amounts of data ~\cite{ZHANG2020100145, hu2024genius, LiFH23}.
Additionally, ML models \cite{de2021machine} achieve satisfactory performance in quantitatively analyzing social science problems across a wide range of popular domains ~\cite{ziems2023large, chenhao2016winning, Zhan2022WhyDY, elsherief2021latent, gabriel-etal-2022-misinfo, baly-etal-2020-detect, ziems-etal-2022-inducing, choi2021meets, demszky2021learning}. 
However, a significant gap remains between the advancements in ML models and their practical deployment in social science research due to their complexity ~\cite{thieme2020machine, grimmer2021machine}.

\minihead{NL2SQL} Natural language access to databases is a key area of research \cite{copestake1990natural}. 
One branch is table question answering systems that understand and answer questions directly based on data presented in tabular form \cite{DBLP:journals/corr/PasupatL15,iyyer-etal-2017-search, zhongSeq2SQL2017, herzig2020tapas, eisenschlos2020understanding}. 
As database sizes and complexity grow, the scalability of NL2SQL systems makes them more viable options ~\cite{ma2023semantic}. 
Existing NL2SQL systems achieve high accuracy ~\cite{2023KatA} by (1) improving the representation of table schema ~\cite{wang2021ratsql, Deng_2021}, and (2) improving the mapping between the intent of natural language queries and the translated SQL codes ~\cite{guo2019complex, yu2018typesql}. 
The generalizability of existing NL2SQL systems has also been widely studied \cite{suhr-etal-2020-exploring} and improved \cite{shaw2020compositional, fu2023catsql}. The development of LLMs makes NL2SQL systems more effective for handling complex database queries 
~\cite{li2024can, pourreza2024din, zhang2023sciencebenchmark}. 
Existing NL2SQL systems and benchmarks addressing vague queries focus on the table schema level \cite{bhaskar2023benchmarking, wang2023know, NL2SQLIA}. 
However, in actual social science research \cite{chenhao2016winning, saravia-etal-2018-carer, Lazer2018The, mohammad-etal-2016-semeval, rothbaum2000cognitive, sharma2020computational, weller-seppi-2019-humor, sap-etal-2020-recollection, zhang-etal-2018-conversations, sap-2022, chu2018learning}, the causes of ambiguity are more complex,
impeding the deployment of NL2SQL systems in social science research ~\cite{qin2022survey}.


\section{Conclusion}
In this work, we present \datasetname, a dataset that comprehensively covers $120$ popular queries in social science research.
Along with \datasetname, we introduce \sysname, an end-to-end library designed to support social science research by automatically analyzing user-collected unstructured data in response to their natural language queries. 
\sysname~ incorporates a forward planning filter that handles vague queries and generates function chains effectively. By integrating innovative structures such as the function tree, doubly linked lists, and alias check blocks, \sysname~ achieves 100\% pass @ 3 and 92\% pass @ 1 with an average end-to-end cost being \$1.06 per query on \datasetname, significantly outperforming the baselines.
\balance

\bibliographystyle{ACM-Reference-Format}
\bibliography{reference}

\end{document}